\documentclass[10pt, a4paper, notitlepage]{article}

\usepackage{tgpagella}  
\usepackage{mathpazo}   

\usepackage{geometry}
\usepackage{url}
\usepackage{amsmath}
\usepackage{balance}
\usepackage{natbib}
\usepackage{graphicx}
\usepackage{booktabs}
\usepackage{tabularx}
\bibpunct[, ]{(}{)}{;}{a}{}{,}

\usepackage[dvipsnames]{xcolor}
\usepackage[colorlinks=true,
  linkcolor=CadetBlue,
  citecolor=CadetBlue,
  urlcolor=CadetBlue]{hyperref}

\usepackage{sectsty}
\sectionfont{\fontsize{12}{12}\selectfont}
       \subsectionfont{\fontsize{10}{10}\selectfont}

\newcounter{quotes}
\newcommand{\source}[1]{$\mathcal{S}_{\refstepcounter{quotes}\label{#1}\ref{#1}}$}
\newcommand{\refsource}[1]{$(\mathcal{S}_{\ref{#1}})$}

\begin{document}

\title{A Critical Correspondence on Humpty Dumpty's Funding for European
  Journalism}

\author{\normalfont Jukka Ruohonen \vspace{4pt}\\ University of Turku, Turku, Finland \\ \url{https://orcid.org/0000-0001-5147-3084}  \\ \texttt{juanruo@utu.fi}}

\maketitle

\begin{abstract}
This short critical correspondence discusses the Digital News Innovation (DNI)
fund orchestrated by Humpty Dumpty---a.k.a.~Google---for helping European
journalism to innovate and renew itself. Based on topic modeling and critical
discourse analysis, the results indicate that the innovative projects mostly
mimic the old business model of Humpty Dumpty. With these results and the
accompanying critical discussion, this correspondence contributes to the ongoing
battle between platforms and media.
\vspace{5pt}
\end{abstract}

\section*{Introduction}

Mass media, including print media and television, has been in a deep crisis ever
since the Web became popular. Although the crisis cannot be attributed to a
single cause, declining advertising revenues, mobile devices, social media, and
the rise of platforms constitute the conventional explanation. In short: a
lion's share of online advertising revenues is captured by a few of
multinational companies, among them Humpty Dumpty.

Scholars of journalism have not shied away in their use of words for describing
the situation. Although not everyone agrees~\citep{Chyi16}, the common narrative
is dreary. Media and journalism have entered into a trap \citep{Myllylahti18} in
which they are in a symbiotic relationship with parasites \citep{Poell20}, the
Internet's true money-making machines, which, however, are compelled to provide
crumbs of funding to media because otherwise there would be no money to be made
from news aggregation, content stealing, and advertising~\citep{Alexander15}. By
2021, it had became also clear that money was not the only issue. Step by step,
the global narrative turned even gloomier. Platforms had become intermediaries
through which democratic deliberation occurs---and, more than often,
poorly~\citep{Nielsen20, Press20}. At the moment, however, after a decade of
wild west, it seems that things are slowly starting to change. Given the
mounting evidence of anti-competitive behavior and other harmful societal
consequences, state intervention is looming throughout the world. While
regulation has been pushed forward particularly by the European Union, antitrust
investigations are underway on both sides of the Atlantic, among other
measures. These recent public policy responses justify the paper's analogy of
Humpty Dumpty, the classical nursery rhyme in which an irreparable downfall
occurs.

As Humpty Dumpty and other platform companies seem to be thus now sitting on a
wall and perhaps fearing for their fall, they have started to implement
different strategies for gaining redemption. Sticks and carrots are both used;
their strategies range from blackmailing (such as threatening to withdraw from a
given country) to monetary aid for those whose assets they exploit. In terms of
the latter strategy, a good example would be the Humpty Dumpty's recent
announcement to pay a billion to news organizations around the
globe~\citep{Reuters20}. The payment was a part of a larger conflict between
media and platforms; another recent battle occurred later between Facebook and
Australia over the former's threat to withdraw from the
country~\citep{BBC21}. By and large, the DNI fund can be seen as a carrot in the
same conflict. It remains to be seen whether these strategies can put
  Humpty Dumpty together again.

Strategies often contain hidden agendas. Against this backdrop, this short
correspondence takes a critical look at the projects funded through the DNI
treasury. What is actually worshiped at the altar of innovation? Is something
sacrificed along the way? Who is sacrificing whom? The rationale for these
critical questions can be justified with an argument that both journalists and
scholars should investigate platforms critically even though they are at the
same time tied to those platforms; there is both a symptom and a
response~\citep{Burgess19, Fitzgerald19}. After a brief further
  theoretical motivation, the materials and tools to examine the symptom and
the response are subsequently elaborated. The critical analysis follows. The
correspondence ends to a few concluding remarks.

\section*{Motivation}

Mass media continues to struggle economically. In essence, Humpty Dumpty and
other platform companies have found a license to print money at the expense of
traditional media. These companies are not publishers, as famously and
repeatedly claimed by the chief executive officer of Facebook---according to
whom the companies are information technology companies instead. Be that as it
may, there is a circular logic present because also traditional mass media
companies are increasingly technology companies themselves. In the new digital
era many companies are cut from the same cloth, the color of which painted with
technology. It is also technology that is seen to provide a solution to the
crisis of mass media.

The solution reflects the enduring vision of technology as a driving force
behind societal change. It is difficult to disagree with the claim. For
instance, many famous models for economic growth assert that technology is an
explanatory factor together with capital, labor, and human capital~\citep[see,
  e.g.,][]{Romer90}. The disagreements start when technology is asserted to be
\textit{the} driving force behind societal change, including economic
growth. Nonetheless, this vision is common in the information technology
industry within which it morphs into business strategies. Whenever there is a
problem, technology is sold as a solution. There are no alternatives, and all
problems are alike; whenever there is a societal problem, technology is again
there to rescue. Because technology drives change, it must be also technology
that provides a gauze for any open wound in a society. Victory has a thousand
fathers but defeat is an orphan; the vision and its logic go under many names,
among these technology exceptionalism~\citep{Doctorow20}, technological
determinism~\citep{Dafoe15}, and technological
solutionism~\citep{Morozov20}. The circular reasoning again starts when a
technology has caused a societal problem, which, given the determinism, needs to
be addressed by a technology, which may cause further problems, which must be
dealt with technological solutions, and so forth. Defeat is an orphan also in
the information technology sector.

Technological determinism has a long history in social sciences, including the
philosophies of science and technology. Usually, but not always, it is embedded
to polemical arguments, the radical argument being that technology and science
are constructed in social settings by engineers and scientists. Such an argument
is as radical as technological determinism itself \citep{Dafoe15}. It is not
necessary to delve into this polemic in detail. For the purposes of this short
critical correspondence, it suffices to point toward \citeauthor{Heidegger77}'s
\citeyearpar{Heidegger77} classical treatise about the philosophy of
technology. Accordingly, to simplify, technology in itself is nothing
technological; the essence of technology is elsewhere. Consequently, technology
cannot help understanding technology philosophically, and, therefore, problems
caused by technology cannot be solved by improving technology or creating new
technologies. On the one hand, this philosophical argument provides a general
theoretical motivation for the critical correspondence; on the other hand, the
argument's social constructivist underpinnings motivate the correspondence's
methodological approach, as soon elaborated in the subsequent section.

To move beyond general assertions about technology, something must be said about
platforms, which are a key ingredient in the current digital economy. There are
many, many platforms powering the digital economy. Paradoxically, however,
platform economy---for a lack of a better term, has always been about
concentration of wealth and power, about winner-takes-all business. One platform
to rule them all; whether it is so-called one-sided platforms, two-sided
platforms, or multi-sided platforms, the core theoretical presumption involves a
critical mass leading to so-called network effects~\citep{Ondrus15,
  Ruohonen18WEIS, Ruutu17}. No one joins a platform with only a few
participants, whether people or companies. Sealed with a kiss: but once a
platform has gained a critical mass, there is a strong incentive for others to
join the bandwagon because the benefits outweigh the potential costs. This logic
is the essence behind the concept of network effects. With respect to Humpty
Dumpty, whose analytics have conquered most of the Internet and whose platform
is the dominant one in mobile phones, there is only a small incentive for a
media company to opt out from the advertising machinery already because most of
the company's competitors are using the same machinery. Given this reasoning,
the DNI projects should also reveal the same basic theoretical components of the
platform economy.

Data powers the platform economy in general and the online advertising business
in particular. Some say data is the new oil, but Humpty Dumpty says it is the
new sunlight, open for anyone to harvest and transform into insights sold for
money \citep{Ghosh20}. Regardless of the particular metaphor one prefers, data
is increasingly also an important element in today's journalism. Analytics and
advertising are the obvious examples, but, in addition, in recent years data
journalism has emerged as a distinct branch of its own. To some extent, data
journalism came into being out of necessity; assessing large data leaks, such as
the so-called Panama papers, necessitated basking in the sunlight of new
technologies and methodologies for journalism. Given this reasoning, again, the
DNI projects should presumably address also data journalism and its
requirements.

Finally, platforms and data together power another kind of an economy. It is
commonly known as the attention economy~\citep{Myllylahti18}. This economy is
based on clicks, eyeballs, and likes---on engagement. Then, what drives
engagement; what makes someone to click and like? The answer has to do with
emotions and emotionality, which are not limited to social media and platform
companies but extend to journalism as well. Fit like a glove: emotionality
sells, and negative emotions and outrage sell even better, particularly in
political journalism and political
advertising~\citep{Ruohonen20MISDOOM}. Emotions are, indeed, particularly
well-suited for political purposes already because emotions are performative;
they are often explicitly used to reach specific goals~\citep{Melacon16,
  WahlJorgensen21}. With respect to politics and political journalism,
\textit{logos} has to some extent lost to \textit{ethos} and particularly
\textit{pathos}. At the same time, negative emotions are widely expressed toward
mass media and journalists by populist political parties and other actors. The
trap to which journalism has entered is not only financial; the crisis of mass
media cannot be explained by economic and technological factors alone
\citep{Alexander15}. Given this reasoning, in any case, also the DNI projects
should uncover something about emotionality, politics, and associated societal
problems.

\section*{Materials and Methods}

The DNI fund boasts to have awarded $150$ million to $662$ media projects in
thirty European countries. (At the same time, Humpty Dumpty's total equity
amounted to about $201$ billion in 2019.) The dataset examined contains the
textual descriptions of all
projects.\footnote{~\url{https://newsinitiative.withgoogle.com/dnifund/dni-projects/}}
These descriptions are short: each project is typically described with a short
summary and a brief overview of the solution proposed.

When combined, however, the amount of textual material is large enough that a
quantitative overview is necessary for helping a qualitative analysis to find
its critical eye. Topic modeling suffices for the task and the latent Dirichlet
allocation (LDA) method for computation. The essence of this probabilistic
method is a decomposition of a corpus into a finite number of latent topics with
a distinct vocabulary~\citep{Blei12}. Each topic is represented by a mixture of
unique terms, and each document is a mixture of the latent topics. In terms of
practical computing, it should be remarked that the LDA method is governed by
two hyperparameters (usually denoted by $\alpha$ and $\beta$), both of which may
affect the results. These must be defined in advance or estimate from
data. Furthermore, LDA applications contain the inherently difficult question
about the number of topics to extract. Also this choice must be made before
interpretation. To aid the decision, the so-called perplexity statistic can be
used as an ``elbow method'' heuristic; lower values are better~(see, e.g.,
\citealt{Griffiths04}). With respect to preprocessing, the following six steps
were taken: (1) each document was tokenized according to white space and
punctuation characters; (2) the tokens were lemmatized into their dictionary
forms and (3) lower-cased; (4) tokens shorter than four characters were excluded
alongside (5) custom stopwords; and (6) only nouns were included via
part-of-speech tagging. The custom stopwords are: \textit{project},
\textit{medium}, \textit{media}, \textit{news}, \textit{article},
\textit{content}, \textit{story}, \textit{journalism}, and
\textit{journalist}. The (term) frequency of the nouns supply the input data to
the LDA method.

The qualitative analysis builds upon the so-called critical discourse analysis
(CDA). As there are no rigorous definitions for the method, it helps to briefly
consider the two main words, discourse and critical. The former takes many
meanings: discourse can refer to a sense-making in a social process, a language
and a vocabulary associated with a particular domain, and as a way for humans to
socially construct something through a language~\citep{Fairclough13}. Although
the last notion is the most contested (see, e.g., \citealt{Hacking99}), it is
also the most interesting one. Discourse may or may not construct reality, but
it sure is a core element in the fundamental concept of power. To suggest
otherwise is to suggest that there would not be power dynamics involved when a
scholar deliberately seeks the right discourse that will convince a review board
to approve a grant application. Birds of a feather flock together: funding,
jobs, positions, and related carrots have been a typical way for companies to
wield power over academia, and platforms are no different in this
regard~\citep{Abdalla21}. The same point applies to journalists and media
companies seeking to fund their ideas via the DNI deposit. Getting a carrot
implies altering one's ideas or at least the discourse describing these. Either
way, one submits to power.

Ergo, the critical part of CDA emphasizes the use of language as a distinct
social practice for exerting, maintaining, and legitimizing power by elites and
powerful institutions, among these mass media~\citep{Bouvier18, vanDijk93},
Humpty Dumpty, and other platform companies. To pin down the CDA method a little
further, the critical focus is typically either a structure or a
strategy~\citep{Fairclough13}. In what follows, the focus is on the latter,
which CDA seeks to reveal by focusing on rhetorical intents, ideological
positions, and presumed audiences~\citep{Henry07}. The dataset suits well the
focus. More than anything, the project descriptions are marketing material, and
with such material rhetorical tricks and ideological flattering are common in
order to impress an audience. Yet the critical inquiry is not meant to ridicule
the projects and journalists involved. In contrast, the critical eye is on the
strategic rhetoric that is required for obtaining media funding from Humpty
Dumpty---the financier and the primary audience, and on how the rhetoric is then
exploited by the company for public relations. If successful, ergo, the short
forthcoming critical discourse analysis should reveal something essential about
today's visions for media and journalism---or at least the visions endorsed by
Humpty Dumpty. If successful, a twinkling of power should be also seen.

\section*{Analysis}

Investigative journalism is usually seen as the crown jewel of journalism. Like
shiny new things, jewels in a crown provide also a convincing persuasion
strategy. The message is clear: investigative journalism needs to reinvent
itself. It needs to innovate. To this end, one project aspires to hold
``\textit{power to account}'' because there ``\textit{are many powerful public
  interest stories out there that will only be discovered if traditional
  investigative techniques are combined with technology}'' \refsource{s: bureau
  local}. This kind of a discourse plays nicely with the perception of many
journalists that data journalism is a norm of ``the twenty-first century, a
technique to help showcase injustice, corruption or inequality''
\citep[p.~10]{Weiss18}. The perception may be true or false, or something
in-between, but the projects examined tend to also send another, implicit
message; that it is not just any technology but \textit{the} technology. The
technology is tied to \textit{the}~infrastructure.

For instance, there is a project wanting to ``\textit{create journalism that
  exposes wrongs in society}'' by journalists who are ``\textit{independent of
  the owners of infrastructure}''~\refsource{s: crji}. But as always, one should
not buy a pig in a poke. It could also be that extensive freelance journalism is
already a problem for the profession---and new \"Uber-journalists are not
necessary the right answer. And it could be that decoupling a newspaper from its
own digital infrastructure is the worst possible business choice that the
newspaper could do. But the trap has been armed in plain sight: among other
things, the DNI giveaway comes with a promotion of Humpty Dumpty's cloud
platform for news organizations. Divide, ignite, conquer, lock in. Along these
steps, it appears that the technology and the infrastructure are further related
to \textit{the} financial pressure. Fortunately, it is possible to build
``\textit{a news source that kept its impartiality and code of ethics intact
  without being susceptible to financial pressures}''~\refsource{s: remp}. The
quantitative LDA results provide a decent panorama for seeing how this hocus
pocus is generally carried out by the DNI projects.

\begin{figure}[th!b]
\centering
\includegraphics[width=10cm, height=4cm]{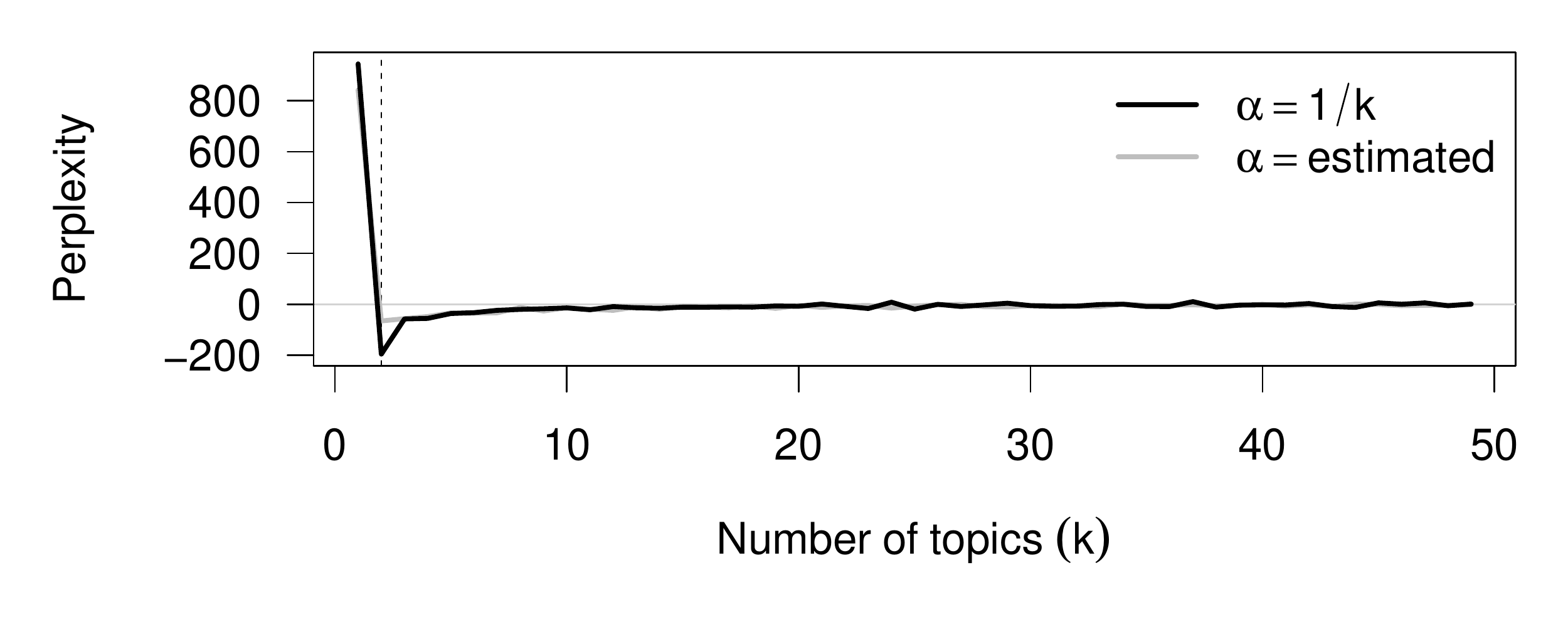}
\caption{Perplexity Values}
\label{fig: perplexity}
\end{figure}

\begin{figure}[th!b]
\centering
\hfill
\minipage{0.50\textwidth}
\includegraphics[width=6cm, height=6cm]{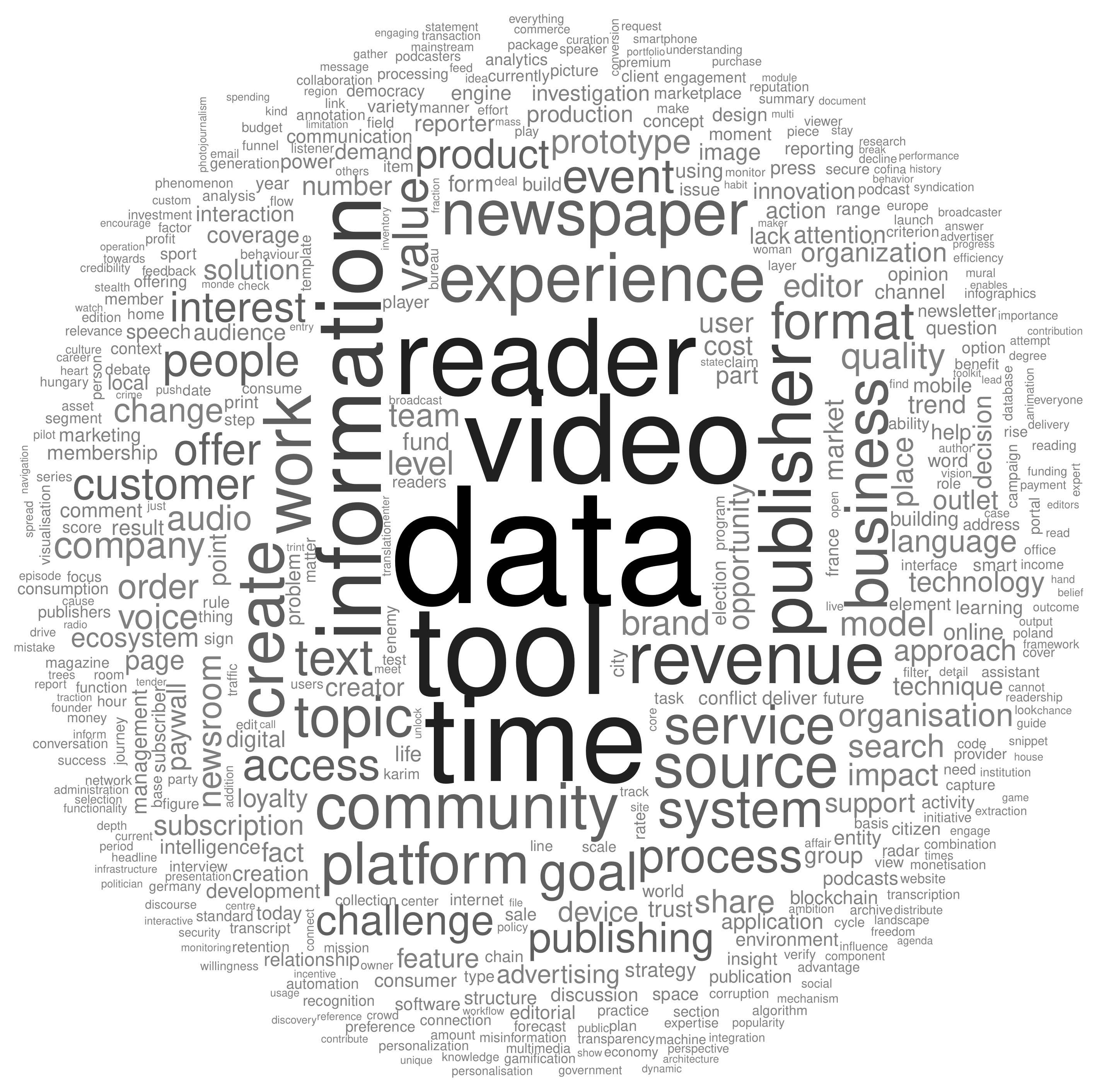}
\caption{The First Topic}
\label{fig: first topic}
\endminipage\hfill
\minipage{0.50\textwidth}
\includegraphics[width=6cm, height=6cm]{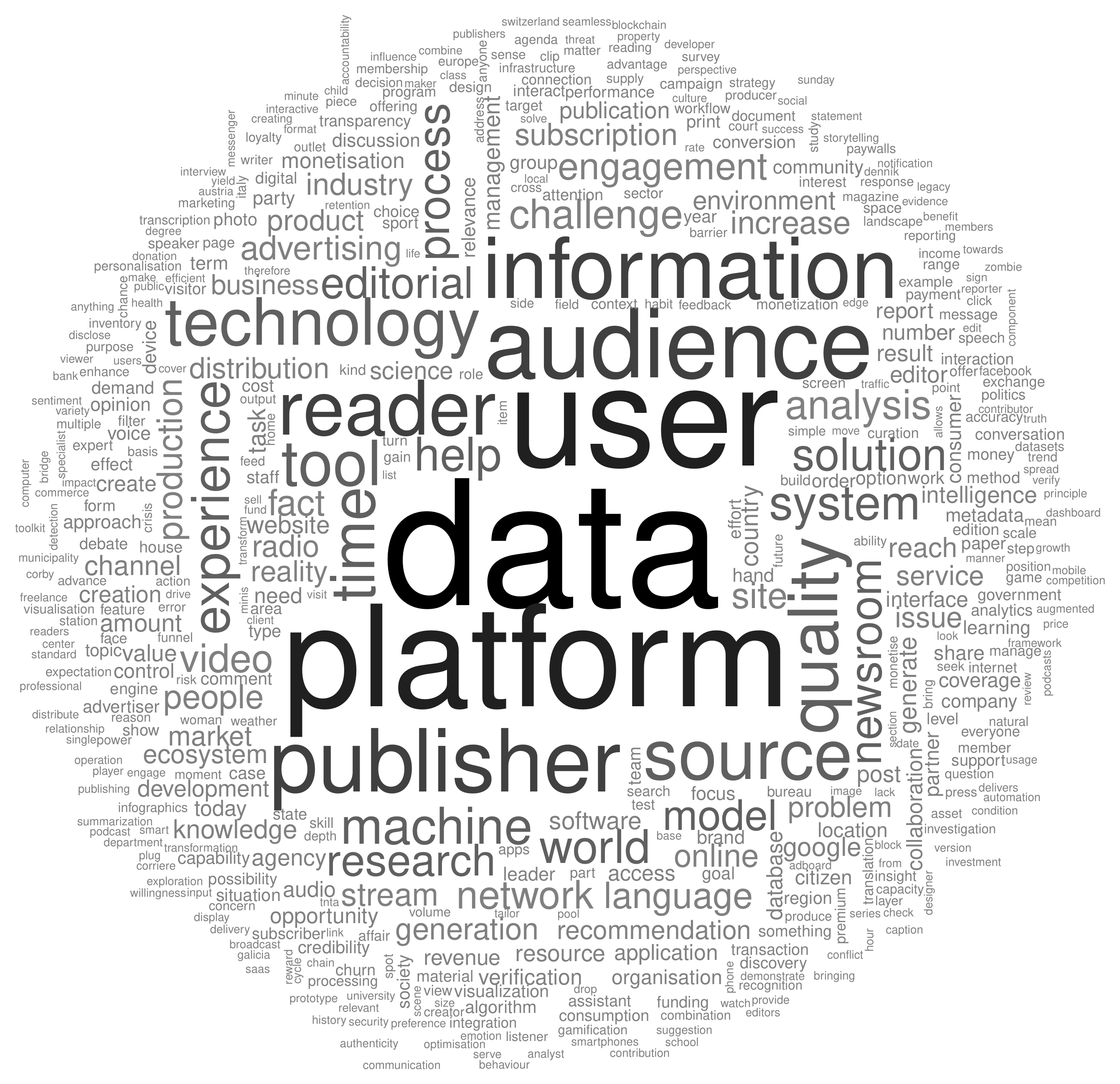}
\caption{The Second Topic}
\label{fig: second topic}
\endminipage\hfill
\end{figure}

Ergo, according to Fig.~\ref{fig: perplexity}, only two topics seem to be enough
for capturing enough of the statistical variation. The result is hardly
surprising because there is only a little variation in the actual substance of
the DNI projects. They are all about technological ``solutions''. See for
yourself: there is one boldfaced noun that stands out from both topics
illustrated in Fig.~\ref{fig: first topic} and Fig.~\ref{fig: second
  topic}. (The larger the text and the darker the font, the higher the
probability that the given noun characterizes the given topic.) Not by accident,
this noun also characterizes Humpty Dumpty's lovely business model. To this end,
the LoVer project acknowledges that ``\textit{programmatic advertising works
  better using data}''~\refsource{s: lover}. A match made in heaven.

But some nuances are still present. On the left-hand side word cloud, the noun
video stands out. By turning to the CDA side, there indeed are interesting new
innovations on the horizon. One of these is the kind of a revolutionary
TikTok-journalism or the ``\textit{Pokemon Go for news}'', implemented via
``\textit{gamification, local and interest-based communities and augmented
  reality contents consumed on mobile phones}''~\refsource{s: newsbub}. On the
right-hand side plot, the word platform is present in big bold font. Again,
hardly surprising as such: what would the new ``15 seconds journalism'' be
without a platform?

The platform projects range from large ecosystems to smaller initiatives. In
terms of the larger endeavors, Humpty Dumpty and other platform companies
``\textit{have raised the bar for frictionless logged-in experiences and the
  news ecosystem needs to keep pace}'' by creating ``\textit{a single-sign-on
  alliance, enhancing the user journey}'' \refsource{s: sso geste}. The lock-in
strategy comes in many disguises. In contrast, the ``\textit{Come Together!}''
project is quite frank: because platforms do not provide journalism, the
solution is simple; the ``\textit{project aims to combine the two things}'' with
the help of ``\textit{a dashboard with widgets}'' \refsource{s: come
  together}. The control panel fetish is quite common, in fact. There are many,
many projects seeking to implement these. In order to sell one, it suffices to
replace the term widget with something else; ``\textit{a real-time
  dashboard}''~\refsource{s: anp}, say. Come Together, allow a Dashboard to be
your Personal God, as you, dear readers, also need personalization so that
engagement and monetization work.

Personalization comes in many forms. For instance, there is something
interesting called ``\textit{personalised hyperlocal couponing}'' \refsource{s:
  sesaab}. But, in general, the likely safest way to obtain funding is to parrot
Humpty Dumpty and other platform companies. To this end, one project seeks
``\textit{to turn the classic, static news website experience into something as
  lively and close to the user as the `personal stream' experience that powers
  social networks}''~\refsource{s: retux}. While it could be debated whether
there is a contradiction between parroting and innovation, for the present
purposes, it is more important to note that the evidence is far from being
straightforward regarding the effects of personalization on the readers'
willingness to subscribe and pay. The lack of effects is noteworthy particularly
for small investigative outlets~\citep{Price17}. It is also important to remark
that the traditional paywall-model is not a polar opposite of the
personalization-based revenue model. Quite the contrary, in fact:
personalization is easier and likely more effective for subscribers whose
personal data can be also correlated with ``anonymous'' visitor
data~\citep{Rubell20}. Although personalization offers some advantages to
consumers, such as potential relevance of the news stories displayed, they are
also increasingly aware of the privacy risks, and feel that social media
advertising, in particular, is mostly irrelevant, annoying, and unethical
\citep{Baglione19, Brinson18}. Given that the use of ad-blocking software is
presumably the largest consumer boycott in human history, a similar point likely
applies to advertising via Humpty Dumpty's personalization ferris wheel.

The preceding quantitative and qualitative points all reflect a traditional
business strategy in the information technology industry: a ball and chain, a
lock-in to a single technical solution, whether a standard, an operating system,
an application programming interface, or a platform. The probably earliest
example of the strategy was seen during the so-called Unix wars in the 1980s and
early 1990s~\citep{Axelrod95, West03}. Thereafter, many similar wars have been
fought, and all is fair in love and war. Though, many of the DNI projects seek
to also address propaganda, polarization, and related ills of contemporary
societies. But it seems that the solutions proposed tend to replicate the root
cause of the ills rather than cure these. One project comes close by focusing on
the ``\textit{the economic revenue from digital advertising}'', yet the solution
proposed merely assigns ``\textit{a score to editorial web pages}'' that allows
``\textit{advertising investors to regain control of the media on which they
  publish}''~\refsource{s: tnta}. Alas, the underlying business model remains
unaltered. Also mass media needs a healing according to the DNI projects.

Ergo, the ``\textit{Constructive Mirror}'' project seeks to improve imbalanced
reporting and negative reporting by infiltrating newsrooms with
``\textit{sentiment analysis}'' that allows ``\textit{news organisations to
  become conscious of their biases}'' \refsource{s: constructive mirror}. The
idea is sound in theory; it is difficult to argue that media would be
unbiased. But an argument that media should be unbiased is a normative stance
and a political statement, even when keeping in mind that media often seeks
balance by covering both sides of an argument \citep{Boulianne20}. Ergo, it
becomes also difficult to agree with the Mirror that mirrors an illusion that
politics and news can and should be free from subjectivity and emotionality;
that media and journalism could and should be trapped into a cage of rationality
and objectivity~\citep{Boler18}. The Mirror further mimicries another common
illusion; that a technological hack would cure societal~ills.

But who is partially responsible for the ills? Could it be that a particular
kind of technology contributes to the ills?  According to the DNI projects, the
answer is negative: the onus is one the readers' side. In fact,
``\textit{readers have their own principles and values as well, often seeking
  for confirmation from like-minded people and news channels; at the risk of
  trapping themselves unintentionally in an echo chamber of their own
  making}''~\refsource{s: belga}. Nowhere is there a project even remotely
suggesting that Humpty Dumpty and other platform companies might be behind the
chambers. So what is a solution then? Well, it is not only newsrooms who need to
be cured by sentiment analysis; a newspaper needs to understand ``\textit{how
  our readers feel}'' via ``\textit{deeper understanding of emotions and their
  role as engagement drivers}''~\refsource{s: keskisuomalainen}. Ergo, holding
``power to account'' is not about holding \textit{the} power to account. It
follows that the Mirror is not, in fact, a mirror but a projection. It is a
projection into what is now known as data journalism, or data science, or data
capitalism~\citep{West17}, or whatever is now prefixed with data. The discourse
backing the projection seemingly seeks to turn news into ``infographics'',
numbers and diagrams, observations without context, plots without plots. Upon
admitting to a viewpoint that ``data speaks for itself'', these become the
facts. Numbers never lie. All biases are foregone; a position that is hard to
maintain even by the most fierce positivist.

The question that follows, then, is what are all these widgets and dashboards,
data and plots, numbers and diagrams, doing to journalism? Thanks to the funding
from Humpty Dumpty, there are projects that provide answers to this question as
well. For one thing, new shiny things are needed so that journalists
``\textit{will be able to prioritise stories based on their forecast
  ROI}''~\refsource{s: prisma} and to measure ``\textit{content performance}''
\refsource{s: anp}. For another thing, data journalism enables
``\textit{journalists to find stories in data more easily and to publish more}''
\refsource{s: crunch}. For many scholars, there may be something eerily familiar
with these answers. For many journalists, however, the answers do not
necessarily ring alarm bells. Metrics already drive editorial decisions in many
newsrooms \citep{Fitzgerald19}. There is only a thin line between love and hate;
viral news benefit both the hosts and the parasites. According to surveys,
furthermore, journalists tend to agree with claims that data science allows to
publish more, improves the quality of news, and opens new avenues for
inquiry---yet without undermining the esteemed journalistic
values~(\citealt{Heravi18}; see also \citealt{Weiss18}). But what the surveys
have not asked is a question about marrying for money; whether
journalists still agree when the data technologies are implicitly or explicitly
tied to the money-making machines and their infrastructures.

\section*{Conclusion}

What is worshiped? Is something sacrificed along the way? Who is sacrificing
whom? The conclusion from the short critical discourse analysis is clear: the
DNI projects are not a reflection but almost a $1:1$ replication of Humpty
Dumpty and its business model. This model is based on data. It is also data and
its analysis that enable the lock-in strategy for the company. Ergo, the first
question and the third question attain their tentative answers. Money comes to
money. The middle question is trickier, but, by argument, data journalism has
not redeemed the pledge promised to it.

The critical analysis also aligns with other recent critical inquiries. Media
and journalism are not only losing their revenues to platforms, but Humpty
Dumpty and other platform companies are increasingly also dictating what kind of
content, publishing strategies, and business models will succeed or
fail~\citep[pp.~79--80]{Press20}, as well as what kind of propaganda will
succeed or fail \citep[p.~160]{Karpf20}. There is an important lesson here:
platforms are not only facilitators for propaganda, but they are also active
propagandists themselves. Another implicit lesson is about CDA, which,
analogously, has been criticized about ideological biases and propagandist
goals, cherry-picking and misinterpreting, and other epistemological and
methodological mischiefs~\citep{Blommaert00, Henry07}. The quantitative LDA
results somewhat downplay this criticism.

Furthermore, critical discourse analysis cannot answer to a question whether
journalists and mass media representatives are blinded by love; whether they
believe the propaganda. As was already remarked earlier, mastering the hype
cycle~\citep{FennRaskino08} is often necessary for obtaining external funding,
whether in academia or in media. Ergo, it could be that the buzzword bingo is
entirely intentional. On one hand, there are also some existing cues perhaps
indicating otherwise.  Because many journalists have neither been educated in
statistics, computer science, or related fields nor have a strong impulse to
learn these, employees trained in other disciplines have often taken the lead in
data-driven journalism~\citep{Nguyen15, Tabary15, Weiss18}. This may also help
to sell shiny things. On the other hand, by 2021, awakening had occurred also in
many newsrooms. Many have speculated that leaving platforms do not necessarily
cost much~\citep{Myllylahti18}. Many see that the LoVer's programmatic
advertising model works poorly. According to the best computer science evidence
available, based on a review of $189$ primary studies, the model is also
basically beyond repair in terms of privacy~\citep{Ullah20}. What was true
yesterday may be untrue today. In other words, the platform dependency may be an
illusion for newspapers~\citep{Meese20}. But what is certain is that the
fistfight between media and platforms will continue also tomorrow.

\section*{Acknowledgements}

This critical correspondence was funded by the Strategic Research Council at the Academy of Finland (grant number~327391).

\bibliographystyle{SageH}

\section*{Appendix}

\begin{table*}[th!b]
\caption{Data Sources for Direct Quotations}
\centering
\label{tab: sources}
\begin{tabularx}{\linewidth}{llX}
\toprule
$\mathcal{S}_i$ & Location \\
\hline
%
%
\source{s: bureau local} & \scriptsize{\url{https://newsinitiative.withgoogle.com/dnifund/report/telling-local-stories/bureau-local-holding-power-account/}} \\
\source{s: crji} & \scriptsize{\url{https://newsinitiative.withgoogle.com/dnifund/insights/liquid-investigations-helping-journalists-collaborate-safely-scale/}} \\
\source{s: remp} & \scriptsize{\url{https://newsinitiative.withgoogle.com/dnifund/insights/remp-dennik-n-creating-sustainable-models-independent-journalism/}} \\
\source{s: lover} & \scriptsize{\url{https://newsinitiative.withgoogle.com/dnifund/dni-projects/lover}} \\
\source{s: newsbub} & \scriptsize{\url{https://newsinitiative.withgoogle.com/dnifund/dni-projects/newsbub/}} \\
\source{s: sso geste} & \scriptsize{\url{https://newsinitiative.withgoogle.com/dnifund/dni-projects/SSO-GESTE-bringing-a-single-ID-to-the-whole-french-digital-media-ecosystem/}} \\
\source{s: come together} & \scriptsize{\url{https://newsinitiative.withgoogle.com/dnifund/dni-projects/come-together/}} \\
\source{s: anp} & \scriptsize{\url{https://newsinitiative.withgoogle.com/dnifund/dni-projects/real-time-usage-and-performance-tracking/}} \\
\source{s: retux} & \scriptsize{\url{https://newsinitiative.withgoogle.com/dnifund/dni-projects/nrc_media/}} \\
\source{s: sesaab} & \scriptsize{\url{https://newsinitiative.withgoogle.com/dnifund/dni-projects/personalised-content-experience/}} \\
\source{s: tnta} & \scriptsize{\url{https://newsinitiative.withgoogle.com/dnifund/dni-projects/TNTA/}} \\
\source{s: constructive mirror} & \scriptsize{\url{https://newsinitiative.withgoogle.com/dnifund/dni-projects/constructive-mirror/}} \\
\source{s: belga} & \scriptsize{\url{https://newsinitiative.withgoogle.com/dnifund/dni-projects/digitally-enable-bias-detection-in-news-articles/}} \\
\source{s: keskisuomalainen} & \scriptsize{\url{https://newsinitiative.withgoogle.com/dnifund/dni-projects/managing-digital-churn-through-understanding-reader-emotions/}} \\
\source{s: prisma} &  \scriptsize{\url{https://newsinitiative.withgoogle.com/dnifund/dni-projects/prisma-orion/}} \\
\source{s: crunch} & \scriptsize{\url{https://newsinitiative.withgoogle.com/dnifund/dni-projects/Crunch-AI-driven-tool/}} \\
\bottomrule
\end{tabularx}
\end{table*}

\end{document}